\documentclass[%
 reprint,
 amsmath,amssymb,
 aps,
 floatfix,
]{revtex4-2}

\usepackage{graphicx}% Include figure files
\usepackage{dcolumn}% Align table columns on decimal point
\usepackage{bm}% bold math
\usepackage{comment}

\begin{document}

%\singlespacing

\title{NMR diffusion in restricted environment approached by a fractional Langevin model}

\author{Felipe Pereira-Alves}
\email{felipepereira@usp.br}
\author{Diogo O. Soares-Pinto}
\email{dosp@ifsc.usp.br}
%\author{T. J. Bonagamba}
%\email{tito@ifsc.usp.br}
\author{Fernando F. Paiva}
\email{fernando.paiva@usp.br}
%\affiliation{Instituto de F\'isica de S\~ao Carlos, Universidade de S\~ao Paulo, CP 369, 13560-970, S\~ao Carlos, S\~ao Paulo, Brazil}
\affiliation{S\~ao Carlos Institute of Physics, University of S\~ao Paulo, PO Box 369, 13560-970, S\~ao Carlos, SP, Brazil}

\begin{abstract}
The diffusion motion of spin-bearing molecules is considerably affected in confined environments. In Nuclear Magnetic Resonance (NMR) experiments, under the presence of magnetic field gradients, this movement can be encoded in spin phase and the NMR signal attenuation due to diffusion can be evaluated. This paper considered this effect in both normal and anomalous diffusion by means of the Langevin stochastic model
%, in terms of the velocity autocorrelation function (VACF), 
within the Gaussian phase approximation (GPA) for a constant gradient temporal profile. The phenomenological approach illustrates the emergence of fractional order exponential decay in the NMR signal and retrieves the classical result from Hanh echo. 
%[\textbf{last update: \today}]
\end{abstract}

\maketitle

\section{Introduction}

Recently much interest has been devoted toward the study of restricted diffusion phenomena. The Nuclear Magnetic Resonance (NMR) techniques, being nondestructive and noninvasive, belongs to the most developed and frequently used tools to study the random motion of nuclear spin-carrying molecules in different confined systems \cite{abragam1961, callaghan1991, callaghan2011, price2009, kimmich2012}. The determination of molecular migration with NMR has a long history, from Hahn's discovery of spin echoes in 1950 \cite{hahn1950} to present-day \cite{grebenkov2007}. A few years after the experimental discovery of NMR by Felix Bloch \cite{bloch1946} and Edward Purcell in 1946 \cite{purcell1946}, Hahn realized that fluctuations in local magnetic fields, as a consequence of spin-bearing molecules diffusion, cause an additional decay in the NMR signal.  Diffusion-based NMR experiments have a wide range of applications from porous media to biomedicine, where measuring molecular diffusion has been used to probe the complex morphology of porous
materials and biological tissues \cite{grebenkov2007}.

Diffusion is a common phenomenon in nature and associated with a variety of processes, resulting in inaccuracy and misleading of the underlying physical phenomena \cite{price1997, price2009}. In this work, the term ``diffusion" will be referred to as the self-diffusion performed by nuclear spin-carrying molecules,
%in the absence of any flow
and can be described as follows: in liquids, under equilibrium conditions with a thermal bath, spin-bearing molecules perform Brownian motion (BM) due to thermal energy, which means that they follow random trajectories, changing their positions, without necessarily the presence of a concentration gradient \cite{stallmach2007, reif2009}. So it can be thought of as a BM in the absence of an applied external force, so that, on average, no displacement is observed; however, molecules that were together, in the same neighborhood initially, will be dispersed as a result of translational motions. On a macroscopic level, this collective behavior, in contrast to microscopic individual movement, exhibits great regularity and follows well-defined dynamic laws. The formulation of this process can be done in a similar way to other diffusion processes, as long as it is possible to establish some distinction in the molecules that perform the self-diffusion \cite{callaghan2011, price2009, grebenkov2007}. As will be seen, diffusion-based NMR experiments provide through magnetic fields gradients a noninvasive way to encode the spin random trajectory by its position.

Classical or normal diffusion occurs when the mean squared displacement (MSD) of the particle during a time interval becomes, for sufficiently long intervals, a linear function of it \cite{chandrasekhar1943, van1992, nelson2020, metzler2000}, 
\begin{eqnarray}
    \langle \textbf{r}^2(t) \rangle \propto t.
\end{eqnarray}
However, in complex environments, the Brownian particle often shows different behaviors, a form of diffusion either slower or faster than normal diffusion. When this linearity breaks down, what is referred to as anomalous diffusion arises and the MSD can be characterized in a power law form \cite{lutz2001}, 
\begin{eqnarray}
    \langle \textbf{r}^2 (t) \rangle \propto t^{\eta}.
\end{eqnarray} 
With exponent $\eta > 1$ we observe superdiffusion; $\eta < 1$, subdiffusion; and $\eta = 1$, normal diffusion is recovered. 

From a mathematical point of view, the phenomenon of normal diffusion is a consequence of the central limit theorem (CLT). From this perspective, anomalous diffusion may be regarded as a situation where the CLT becomes inapplicable for one reason or another. One of the possible explanations for the emergence of this behavior is if the translational displacements are restricted by geometrical constraints (walls, membranes, etc.), or the space in which diffusion takes place is of a fractal nature \cite{metzler2000}. 
As a consequence, deviations from Gaussian distribution functions destroy the Markovian character from normal (unrestricted) diffusion. 
These deviations can be taken into account by autocorrelation functions, where memory effects are manifested \cite{kubo1985}. Here, we follow this description by means of the Langevin approach \cite{langevin1908, coffey2004, grebenkov2013, grebenkov2014}. 

This work explores the Hahn echo experiment, where the spins are free to diffuse in some restricted environment and in the presence of a constant gradient. The well-known formula for the signal attenuation $E$ due to diffusion \cite{hahn1950},
\begin{equation}
    E = 
    \exp \left\{-
    \frac{1}{12} \gamma_{n}^2 g^2 D t^3
    \right\},
\label{eq: hahn echo}
\end{equation}
describes the simplest case of unrestricted diffusion, where we emphasize here that the echo amplitude decay as the cube of the echo time, $t^3$. The Eq.(\ref{eq: hahn echo}) is also often used, e.g., when considering bounded regions arguing that the spins do not see the boundaries in the limit of short enough times \cite{grebenkov2007}. The echo amplitude also decays as the square of the applied gradient, $g^2$. The validity of this result is based on the so-called Gaussian phase approximation (GPA) for the spin phase, and it applies for small gradients where the dephasing is small \cite{callaghan2011, price2009, grebenkov2007}. Here, $\gamma_{n}$ is the nuclear gyromagnetic ratio and $D$ stands for the diffusion coefficient. 
Further development of this result has been made to include, for instance, geometrical restriction of diffusing nuclei or temporal dependence and spatial inhomogeneity of the magnetic field \cite{grebenkov2007}. 

Following the ideas above, this paper aims for two purposes. To introduce a constant magnetic field gradient to encode spins precession frequency from their position and, once spatially distinguishable, to apply Langevin's phenomenological approach to model their stochastic position due to diffusion, mainly by its velocity autocorrelation
function (VACF). Two special cases will be considered, namely, normal and anomalous diffusion. As already mentioned, the former is associated with the unrestricted diffusion case, while the latter reveals the complex diffusion motion experienced by the nuclei, e.g., in the presence of geometric restrictions. A useful approach to anomalous
diffusion is the generalized Langevin equation (GLE) where the friction term contains an integral expression describing the intrinsic memory of the environment and it depends upon the system under consideration. We will provide a description in terms of a memory kernel that decays as a power law based on systems that unveil subdiffusion \cite{grebenkov2013, grebenkov2014, cooke2009}. In doing so, we will extend and retrieves the classic result from Hanh echo given by Eq.(\ref{eq: hahn echo}) \cite{lisy2016attenuation, lisy2017attenuation, lisy2018nmr}.

The paper is organized as follows. Section \ref{sec: NMR diffusion} introduces a concise description of the theory behind diffusion-based NMR experiments: magnetic field gradients (Sec. \ref{subsec: Diffusion-weighting magnetic field}), Gaussian phase approximation (Sec. \ref{subsec: Gaussian phase approximation}), and pulse sequence (Sec. \ref{subsec: Pulse sequence}); Sec. \ref{sec: Langevin model} is devoted to our phenomenological model and its features; In Sec. \ref{sec: Explicit Solution} we report our results, where two special cases will be considered: anomalous diffusion (Sec. \ref{subsec: Anomalous diffusion}) and normal diffusion (Sec. \ref{subsec: Normal diffusion}). Then in Sec. \ref{subsec: NMR signal attenuation} they are applied to calculate the NMR signal attenuation.
At last, in Sec. \ref{sec: Conclusions Remarks} we address some concluding remarks in our study.

\section{NMR diffusion}
\label{sec: NMR diffusion}
\subsection{Diffusion-weighting magnetic field}
\label{subsec: Diffusion-weighting magnetic field}
To establish a distinction between the molecules that performs the self-diffusion and make the study of diffusion possible, a magnetic field gradient is applied, also known as diffusion-weighting magnetic field, responsible for labelling the nucleus by changing their precession frequency according to their position. The simplest and most common choice is a spacial constant gradient, so at a point shifted from the initial position by $\textbf{r}$, the inhomogeneous magnetic field $\textbf{B}_G(\textbf{r},t)$ varies linearly in space. Hence, we define the local magnetic field as
\begin{eqnarray}
\textbf{B}(\textbf{r},t) = \textbf{B}_0 +  \mathcal{G}(t) \cdot \textbf{r}, 
\label{eq: gradiente tensor}
\end{eqnarray}
with $\mathcal{G}$ being a tensor, in order to  satisfy the Maxwell's equations \cite{callaghan1996}. The Eq.(\ref{eq: gradiente tensor}) has two contributions: the first one is a strong static magnetic field $\textbf{B}_0$, responsible for inducing equilibrium magnetization; the second one is a linearly varying field $\mathcal{G}(t) \cdot \textbf{r} $, responsible for encoding the nucleus by its position. 
%The first affects all the nuclei, in the same way, given by the Larmor frequency $\omega_0 = \gamma_{n} B_0$; the second is needed to distinguish the precession from its position. 

In practice, one usually has the condition of weak gradients \cite{callaghan1996, price2009, callaghan2011}, $|\mathcal{G}(t)| << |B_0|\hat{\textbf{z}}$, so the components of the gradient perpendicular to the direction defined by the external field $\textbf{B}_0$, that we will consider to be the $z$-axis, can be neglected. The remaining components of the gradient tensor, denoted by $\textbf{g}(t)$, are those parallel to the direction defined by $\textbf{B}_0$. So, from Eq.(\ref{eq: gradiente tensor}), the external field takes the form
%The magnetic field $\textbf{g}(t)$, though inhomogeneous, has axial symmetry.
\begin{eqnarray}
B_z(\textbf{r},t) = \textbf{B}_0 +  \textbf{g}(t) \cdot \textbf{r} .     
\label{eq: campo com gradiente}
\end{eqnarray}  
The Eq.(\ref{eq: campo com gradiente}), despite providing the necessary spatial distinction for the study of diffusion, does not consider the molecules stochastic motion, which causes fluctuations in the position $\textbf{r}$. When the position is treated as a stochastic process $\textbf{r}(t)$, each nucleus acquires a random phase $\Phi(\textbf{r}(t),t)$ due to inhomogeneous magnetic field, which is obtained by integrating the position-dependent precession frequency along the random trajectory $\textbf{r}(t)$ of the nucleus,
\begin{eqnarray}
    \Phi(\textbf{r}(t),t) = \gamma_{n} \int_{0}^{t} \textbf{g}(t') \cdot \textbf{r}(t')\mathrm{d} t'.
\label{eq: fase aleatoria}
\end{eqnarray}
Where Eq.(\ref{eq: fase aleatoria}) is written in a rotating reference frame with the constant Larmor frequency $\omega_{0} = \gamma_{n} B_{0}$, which has no particular interest in the measurement of motion. The random variable $\Phi(\textbf{r}(t),t)$ is now a functional of the stochastic process $\textbf{r}(t)$, with initial phase $\Phi(\textbf{r},0) = 0$, which expresses the condition of isochromatic spins (in phase) right after the $\pi/2$ RF-pulse (see \ref{subsec: Pulse sequence}). 

Thus, from Eq.(\ref{eq: fase aleatoria}) each nucleus contributes to the signal attenuation by the factor $e^{i\Phi}$. However, most NMR experiments are performed with a large number of nuclear spins and the detected signal arises not from fluctuations of individual spins but rather from a coherent superposition over all of them. The theoretical approach to this problem is based entirely on the probabilistic interpretation of this superposition \cite{callaghan1996, grebenkov2007}.
In doing so, the contribution from diffusion to the macroscopic signal can be obtained by averaging the functional $e^{i\Phi}$ over all the nuclei. Since the number of nuclei is very large, the average can be replaced by the expectation over all realizations of the random variable $\textbf{r}(t)$ \cite{grebenkov2007},
\begin{eqnarray}
    E = \langle e^{i \Phi} \rangle.
\label{eq: sinal difusao}
\end{eqnarray}
Where the signal in Eq.(\ref{eq: sinal difusao}) is normalized to one in the absence of diffusion-weighting gradient and when no relaxation is present. Note that the signal attenuation due to diffusion in Eq.(\ref{eq: sinal difusao}) takes the form of the characteristic function of the random variable phase $\Phi$ and leads to an attenuation of the echo due to phase incoherence arising from the stochastic BM. 

\subsection{Gaussian phase approximation}
\label{subsec: Gaussian phase approximation}

From the statistical point of view, the accumulated phase is a random variable whose distribution is defined by several factors, such as the diffusive motion, the applied magnetic field and the geometry of the confining medium \cite{price2009, callaghan2011, grebenkov2007}.
From the cumulant expansion for this variable in Eq.(\ref{eq: sinal difusao}), it is straightforward to see that for antisymmetric gradient profiles with respect to time, all odd-order moments vanish, and only the even-order terms remain (see \ref{subsec: Pulse sequence}). The framework of the Gaussian approximation arises when one neglects higher-order terms in the remaining terms of the cumulant expansion, that are expected to be small at weak gradients regime since from Eq.(\ref{eq: fase aleatoria}) the phase variable $\Phi$ is proportional to the gradient $\textbf{g}$. So the lowest order term contribution is given by the second moment. From this, the Eq.(\ref{eq: sinal difusao}) reduces to 
\begin{eqnarray}
    E = e^{- \frac{1}{2} \langle\Phi^2\rangle}.
\label{eq: atenuacao da fase gaussiana}
\end{eqnarray}
In this framework, averaging over the variance in Eq.(\ref{eq: atenuacao da fase gaussiana}) rather than the exponent in Eq.(\ref{eq: sinal difusao}) is a substantially less challenging problem \cite{grebenkov2007}.

One can work with the spin velocities rather than positions. To do so, we perform an integration by parts in Eq.(\ref{eq: fase aleatoria}) and get
\begin{eqnarray} 
    \Phi(\textbf{r}(t),t) = - \gamma_{n} \int_{0}^{t} \dot{\textbf{r}}\left(t_{1}\right) \left[ \int_{0}^{t_{1}} \textbf{g}\left(t'\right) \mathrm{d} t' \right] \mathrm{d} t_{1},    
\label{eq: integracao partes}
\end{eqnarray}
where the first term $\gamma_{n} \textbf{r}(t) \int_{0}^{t} \textbf{g}(t')\mathrm{d} t' $ vanishes if we consider that the total accumulated phase of immobile nuclei would be zero when integrated up to the signal acquisition time, i.e.,
\begin{equation}
    \int_{0}^{T_E} \textbf{g}(t')\mathrm{d} t'= 0. 
\label{eq: refocalizacao}
\end{equation}
Here, $T_E$ denotes the echo time (see \ref{subsec: Pulse sequence}). This fact represents the so-called rephasing condition of the gradient and is satisfied by all pulse sequences that produce an echo, 
in the absence of any background gradient (no susceptibility effect, perfect shimming of the system, etc.). So Eq.(\ref{eq: integracao partes}) reduces to
\begin{eqnarray}
    \Phi(\textbf{r}(t),t) = -\gamma_{n} \int_{0}^{t} \textbf{G}(t') \dot{\textbf{r}}\left(t'\right)  \mathrm{d} t'. 
\label{eq: fase aleatoria final}
\end{eqnarray}
With $\textbf{G}(t)=\int \textbf{g}\left(t'\right) \mathrm{d} t'$. Now the spins phase depends upon its velocity $\dot{\textbf{r}}$. Back in Eq.(\ref{eq: atenuacao da fase gaussiana}), the second moment can be written in terms of the velocity autocorrelation function (VACF),
\begin{eqnarray}
\left\langle \frac{\Phi^{2}}{2}\right\rangle &&=  
\gamma_{n}^{2}   \nonumber \\ 
&&
%\times
\int_{0}^{T_E} 
\int_{0}^{T_E} 
t_{1} t_{2}
\textbf{G}(t_1) 
\textbf{G}(t_2) 
\left\langle\dot{\textbf{r}}(t_{1}) \dot{\textbf{r}}(t_{2})\right\rangle
\mathrm{d} t_{1} 
\mathrm{d} t_{2}.~~~~~
\label{eq: fase gaussiana}
\end{eqnarray}
%or,
%\begin{eqnarray}
%\left\langle \frac{\Phi^{2}}{2}\right\rangle =
%\int_{0}^{T_E} \mathrm{d} t_{1} \textbf{G}(t_1) 
%\int_{t_{1}}^{T_E} \mathrm{d} t_{2} 
%\textbf{G}(t_2)   \left\langle\dot{\textbf{r}}(t_{1}) \dot{\textbf{r}}(t_{2})\right\rangle.~~~~
%\label{eq: fase gaussiana}
%\end{eqnarray}
%Where we used the symmetry of the VACF to order time variables $t_1$ and $t_2$.
%Where we used the direct analysis of the integration domain to order time variables $t_1$ and $t_2$.
%Here, $T_E$ denotes the echo time (see \ref{subsec: Pulse sequence}). 
Therefore, for the calculation of Eq.(\ref{eq: fase gaussiana}), that is, to evaluate the attenuation of the NMR signal whitin the GPA in Eq.(\ref{eq: atenuacao da fase gaussiana}), it is enough to know the gradient temporal profile and the VACF. 
%In this paper, we shall restrict ourselves to a constant gradient profile and the VACF will be determined by means of the Langevin model. 

\subsection{Pulse sequence}
\label{subsec: Pulse sequence}
\begin{figure}
\includegraphics[width=8cm,height=10cm]{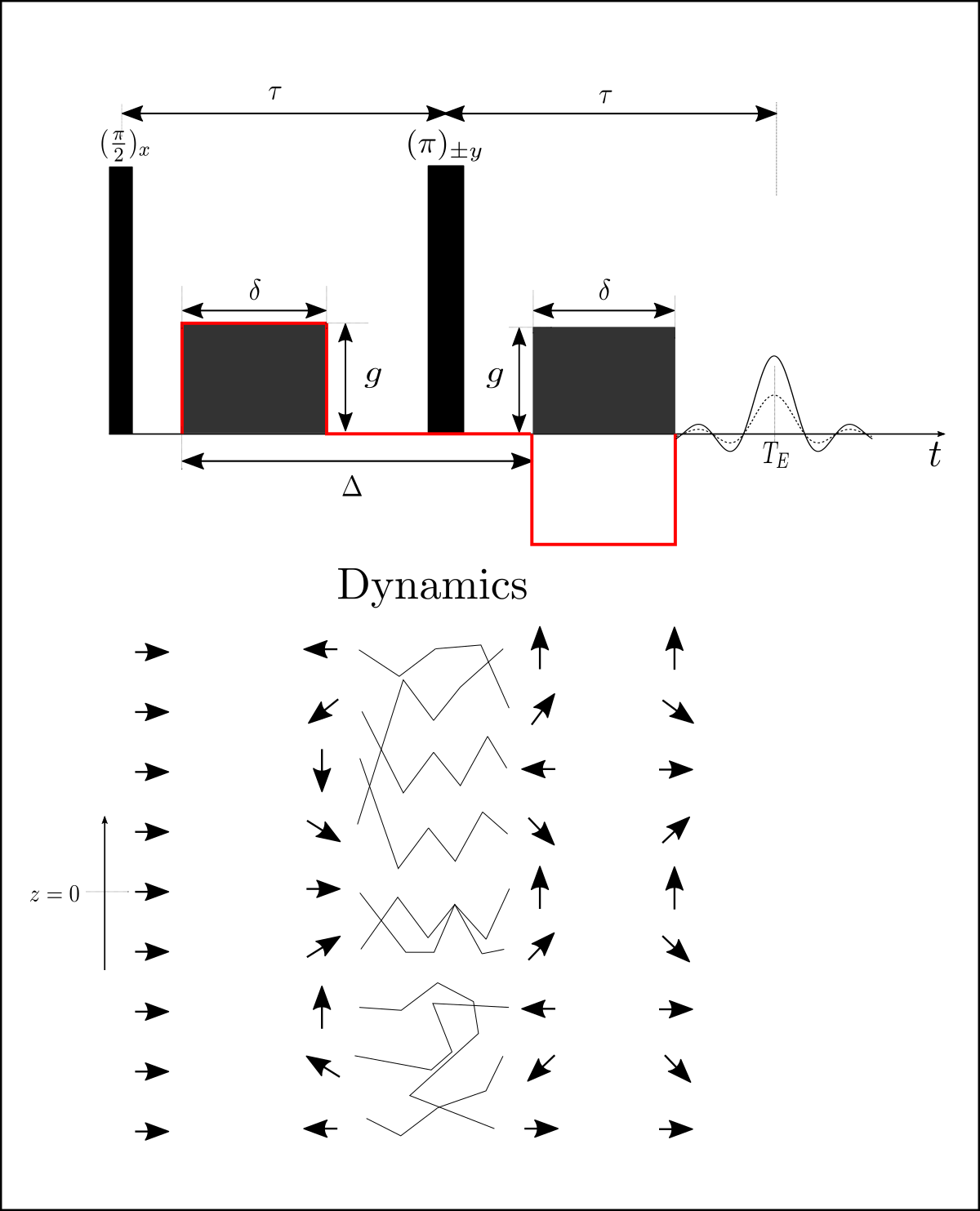}
\caption{Schematic illustration for the pulse sequence described in Sec. (\ref{subsec: Pulse sequence}) when seen in a rotating reference frame with Larmor frequency. 
Here, the thick red line represents the effective temporal profile of the gradients with amplitude $g$, in which the second gradient polarity is inverted by the $\pi$ RF-pulse; $\delta$, the duration of the gradients; $\Delta$, the  gradients spacing. The echo emerges at $T_E=2\tau$,
%Note that the $\pi$ RF-pulse inverts the phase and a subsequent gradient pulse of the same sign has the effect of rewinding. 
when all the spin isochromats come into phase (continuous black line),
where $\tau$ is the inter-pulse delay.
So any translational motion due to diffusion will cause perturbations to the phase evolution and add an attenuation in the NMR signal (dashed black line).}
\label{fig: fig1}
\end{figure}
Diffusion-weighted NMR sequences are made sensitive to diffusion by the addition of magnetic field gradients \cite{carr1954, torrey1956, stejskal1965}. 
%The simplest conceptual method to verify the translational molecular self-diffusion is by means of a steady magnetic-field gradient, which can be briefly described as follows. 
Let us consider the diffusive motion of a nucleus in a given pulse sequence, described as follows (see Fig.\ref{fig: fig1}). 
Under a static magnetic field, present throughout all the experiment, the equilibrium longitudinal magnetization is generated. At time $t = 0$, an application of a $\pi/2$ RF-pulse gives rise to the transverse magnetization, followed by the formation of the NMR signal, when this disturbance ceases. Then, three subsequent intervals of distinct times can be classified for the diffusion-weighting magnetic field: 
after signal generation, the gradient is applied during the period $\delta$, which produces an almost instantaneous phase shift for the nucleus, depending on its position (encoding period). After the spatial distinction, they can diffuse in a time interval denoted by $\Delta$ (diffusing interval), which is the time interval that separates the beginning of gradient fields. This period is responsible for the evolution of the spin phase. Finally, a second gradient is applied also with duration $\delta$
%with opposite polarity to the first one 
(decoding period). If any spin has performed some displacement during the time $\Delta$, a partial phase recovery and an attenuation in the signal will be detected.

The application of a $\pi$ RF-pulse can be taken into account through an effective gradient profile $\textbf{g}^{*}(t)$ for which the magnetization direction is inverted after $\tau$. Since gradients before and after $\pi$ RF-pulse are supposed to be identical, the effective gradient profile $\textbf{g}^{*}(t)$ is antisymmetric with respect to the time $\tau$, i.e.,
\begin{eqnarray}
    \mathbf{g^{*}}(\tau-t)=-\mathbf{g^{*}}(t).
\end{eqnarray}
In particular, the rephasing condition in Eq.(\ref{eq: refocalizacao}) required for echo formation is automatically satisfied. From now on, we stress that the effective gradient profile will be referred to only as $\textbf{g}(t)$.
%and the RF pulses are assumed to be very short, so one can neglect their durations.

The possibility of controlling the temporal profile of the gradient field, as well as its amplitude and direction, allows one to obtain different information from the system. The temporal profile is typically trapezoidal in an experiment, but for simplicity in theoretical analysis the rectangular form is assumed with an amplitude given by $|\textbf{g}(t)| = g$, where the simplest case consists of a steady gradient, $\delta = \Delta = \tau$. 
Finally, one must set the direction to measure the diffusion, denoted by the verse $\hat{\textbf{e}}$, that will be considered fixed.
%, $\textbf{g}(t) = g\hat{\textbf{e}}$. 
So the effective gradient is
\begin{eqnarray}
    \textbf{g}(t) = 
    \begin{cases}
    \textbf{g}, & 0 \leqslant t \leqslant \tau \\ 
    %0, & t=\tau \\ 
    -\textbf{g}, & \tau < t \leqslant 2\tau
    \end{cases}
\label{eq: gradiente efetivo}
\end{eqnarray} 
Imposing these considerations in Eq.(\ref{eq: fase gaussiana}), we get
%\begin{eqnarray}
%\left\langle \frac{\Phi^{2}}{2}\right\rangle = 
%\gamma_{n}^{2}g^{2} \int_{0}^{T_E} \mathrm{d} t_{1} t_{1} 
%\int_{t_{1}}^{T_E} \mathrm{d} t_{2} t_{2}  \left\langle\dot{X}(t_{1}) \dot{X}(t_{2})\right\rangle.~~~
%\label{eq: fase gaussiana numa direcao}
%%\end{eqnarray}
%\begin{eqnarray}
%\left\langle\Phi^{2}/2\right\rangle = 
%\gamma_{n}^{2}g^{2} 
%\int_{0}^{T_{E}} \mathrm{d} t_{2} t_2 
%\int_{0}^{t_{2}} \mathrm{d} t_{1}  t_1 
%\left\langle\dot{X}(t_{1}) \dot{X}(t_{2})\right\rangle.~~~~
%\label{eq: fase gaussiana numa direcao}
%\end{eqnarray}~
\begin{eqnarray}
\left\langle \frac{\Phi^{2}}{2}\right\rangle = 
\gamma_{n}^{2}g^{2} 
\int_{0}^{T_E} 
\int_{0}^{T_E} 
t_{1} t_{2}
\left\langle\dot{X}(t_{1}) \dot{X}(t_{2})\right\rangle
\mathrm{d} t_{1} 
\mathrm{d} t_{2}.~~~~
\label{eq: fase gaussiana numa direcao}
\end{eqnarray}
Where  $\hat{\textbf{e}} \cdot \dot{\textbf{r}} = \dot{X}$, so only the correlation between velocity components along the direction $\hat{\textbf{e}}$ of applied gradients takes part. Here, $T_E=2\tau$.

%\section{Phenomenological model}
\section{Langevin model}
\label{sec: Langevin model}

To describe the temporal evolution of a tracer of mass $m$ in contact with a thermal bath of temperature $T$, we will consider a phenomenological model based on a linear generalization of the Langevin equation (GLE) which reads like Newton's law \cite{van1992, grebenkov2013, grebenkov2014},
\begin{eqnarray}
    m \ddot{X}(t) = F_E(t) + F_S(t) + F(t).
\label{eq: lang. geral 1}
\end{eqnarray}
Here, $X=X(t)$ is the displacement of the tracer along a coordinate and $\dot{X}(t) = V(t)$ refers to velocity. The right side of Eq.(\ref{eq: lang. geral 1}) expresses the balance of forces acting on the tracer:

(i) $F_E(t)$ is the external force, which can represent, for example, magnetic, electric fields, Hookean force, etc.

(ii) $F_S(t)$ is the generalized Stokes force, with a friction given by a memory kernel $\gamma(t)$, and represents the viscoelastic properties of the medium,
\begin{eqnarray}
    F_S(t) = - \int_{-\infty}^{t} \gamma(t-t') \dot{X}(t')~ \mathrm{d}t'.
\label{eq: stokes com memoria}
\end{eqnarray}
For the interpretation of experimental trajectories, the initial time can be fixed as $t=0$, so that the causality principle allows one to cut the integrals below 0 in Eq.(\ref{eq: stokes com memoria}). As a consequence, and due to the linearity of the GLE, standard Laplace transform techniques can be used for a formal solution to the problem.

(iii) $F(t)$ is the rapidly fluctuating thermal force (noise) resulting from molecular collisions. This is the residual force exerted by neighboring molecules - or thermal bath - when the frictional force $F_S(t)$ is subtracted. %The rapid fluctuations, represented by $F(t)$, around the mean value, $F_S(t)$, guarantee the permanent characteristic of Brownian motion XX.

The thermal force $F(t)$ is a stationary and Gaussian noise. The latter condition implies that its distribution is fully characterized by its mean, which we will consider for simplicity as a zero-mean $\langle F(t) \rangle$ = 0,
%if this is not so, we can redefine F to include 
and the covariance $\langle F(t) F(t') \rangle$. In the case of internal noise, the memory kernel is related to the noise correlation function via the fluctuation-dissipation theorem \cite{kubo1985},
\begin{eqnarray}
    \langle F(t) F(t') \rangle = k_B T \gamma(|t-t'|), 
\label{eq: 2 teorema}    
\end{eqnarray}
with $\langle \cdots \rangle$ denoting the ensemble average over random realizations of the thermal force $F(t)$, $k_B$ being the Boltzmann constant and $T$ the absolute temperature. In this case, the noise and dissipation have the same origin and the system will reach the equilibrium state \cite{kubo1985}.
%, which is not the case for external noise \cite{vinales2006, desposito2009}. 

In what follows, we will consider $m=1$ and an isolated system, such that $F_E(t) = 0$, for $t \geq 0$. Combining the above equations, the GLE is given by
\begin{eqnarray}
    \ddot{X}(t) + \int_{0}^{t} \gamma(t-t') \dot{X}(t')~\mathrm{d}t' = F(t), \quad t \geq 0.
\label{eq: lang. geral 2}
\end{eqnarray}
Other forces can be added in Eq.(\ref{eq: lang. geral 2}), however, it describes the dynamics of a large class of problems, depending on the memory kernel $\gamma(t)$. From now on, since we only deal with times $t \geq 0$, we omit the modulus in the argument of functions.
%In what follows, we will consider a system in the presence of two memory kernels.

\section{Explicit Solutions}
\label{sec: Explicit Solution}

By means of the Laplace transform, one can obtain a formal expression for the displacement $X(t)$ and the velocity $V(t)$ expressed by the GLE in Eq.(\ref{eq: lang. geral 2}). So, in the Laplace space, we have \cite{vinales2006, vinales2007, desposito2009}
\begin{eqnarray}
\hat{X}(s) &=& \frac{x_{0}}{s}+v_{0} \hat{G}(s)+\hat{G}(s) \hat{F}(s), 
\label{eq: trans x GLE} \\
\hat{V}(s) &=& v_{0} \hat{g}(s)+\hat{g}(s) \hat{F}(s),
\label{eq: trans v GLE} 
\end{eqnarray}
where $\mathcal{L}[F(t)]=\hat{F}(s)$, $\mathcal{L}[X(t)]=\hat{X}(s)$ and $\mathcal{L}[V(t)]=\hat{V}(s)$ gives the Laplace transform of this quantities. Here, $t$ and $s$ are, respectively, the variables in the time domain and Laplace domain. The functions $G(t)$ and $g(t)$ are defined as, respectively, 
\begin{eqnarray}
\hat{g}(s)&=&\frac{1}{s+\hat{\gamma}(s)}, \label{eq: g trans} \\
\hat{G}(s)&=&\frac{s^{-1}}{s+\hat{\gamma}(s)}, \label{eq: G trans}
\end{eqnarray}
and represents the relaxation functions. 
Let's define another function as
\begin{eqnarray}
\hat{I}(s)&=&\frac{s^{-2}}{s+\hat{\gamma}(s)}, \label{eq: I trans}
\end{eqnarray}
that will be used later to analyze the displacement correlation function.

The case of thermal initial conditions are considered,
\begin{eqnarray}
    x_0 = 0, \,\,\,\,\,\, v_0^2 =  k_B T,%/m.
\label{eq: termal}
\end{eqnarray}
where $X(0) = x_0$ and $V(0) = v_0$. Furthermore, for the memory kernel the following assumption should be satisfied
\begin{eqnarray}
\lim_{t \to \infty} \gamma(t)=\lim _{s \to 0} s \hat{\gamma}(s)=0,
\end{eqnarray}
where $\mathcal{L}[\gamma(t)](s) = \hat{\gamma}(s)$ is the Laplace transform of $\gamma(t)$. 

By applying inverse Laplace transform to Eqs.(\ref{eq: trans x GLE}) and (\ref{eq: trans v GLE}), it follows,
\begin{eqnarray}
X(t) &=& \langle X(t)\rangle+\int_{0}^{t} G\left(t-t'\right) F\left(t'\right) \mathrm{d} t', 
\label{eq: x GLE} \\
V(t) &=& \langle V(t)\rangle+\int_{0}^{t} g\left(t-t'\right) F\left(t'\right) \mathrm{d} t',
\label{eq: v GLE} 
\end{eqnarray}
where,
\begin{eqnarray}
\langle X(t)\rangle &=& x_{0}+v_{0} G(t), \\
\langle V(t)\rangle &=& v_{0} g(t).
\end{eqnarray}
From the behavior of Eqs. (\ref{eq: g trans})-(\ref{eq: I trans}) and its asymptotic limits, one can reveal the diffusion motion features of the particle in a given system by the MSD, time dependent diffusion coefficient and VACF, in the following way \cite{vinales2006, vinales2007, desposito2009}, 
\begin{eqnarray}
\left\langle X^{2}(t)\right\rangle &=&  2 k_{B} T I(t),
\label{eq: msd geral}\\
D(t) &=&  \frac{1}{2} \frac{\mathrm{d}}{\mathrm{d} t}\left\langle X^{2}(t)\right\rangle=k_{B} T G(t), 
\label{eq: D geral}\\
C_{V}(t) &=&  \langle V(t) V(0)\rangle = \left\langle V^{2}(0)\right\rangle g(t).
\label{eq: vacf geral}
\end{eqnarray}
%In what follows, we will consider a system in the presence of two memory kernels.

\subsection{Anomalous diffusion}
\label{subsec: Anomalous diffusion}

\begin{figure*}
\includegraphics[width=18cm,height=12.5cm]{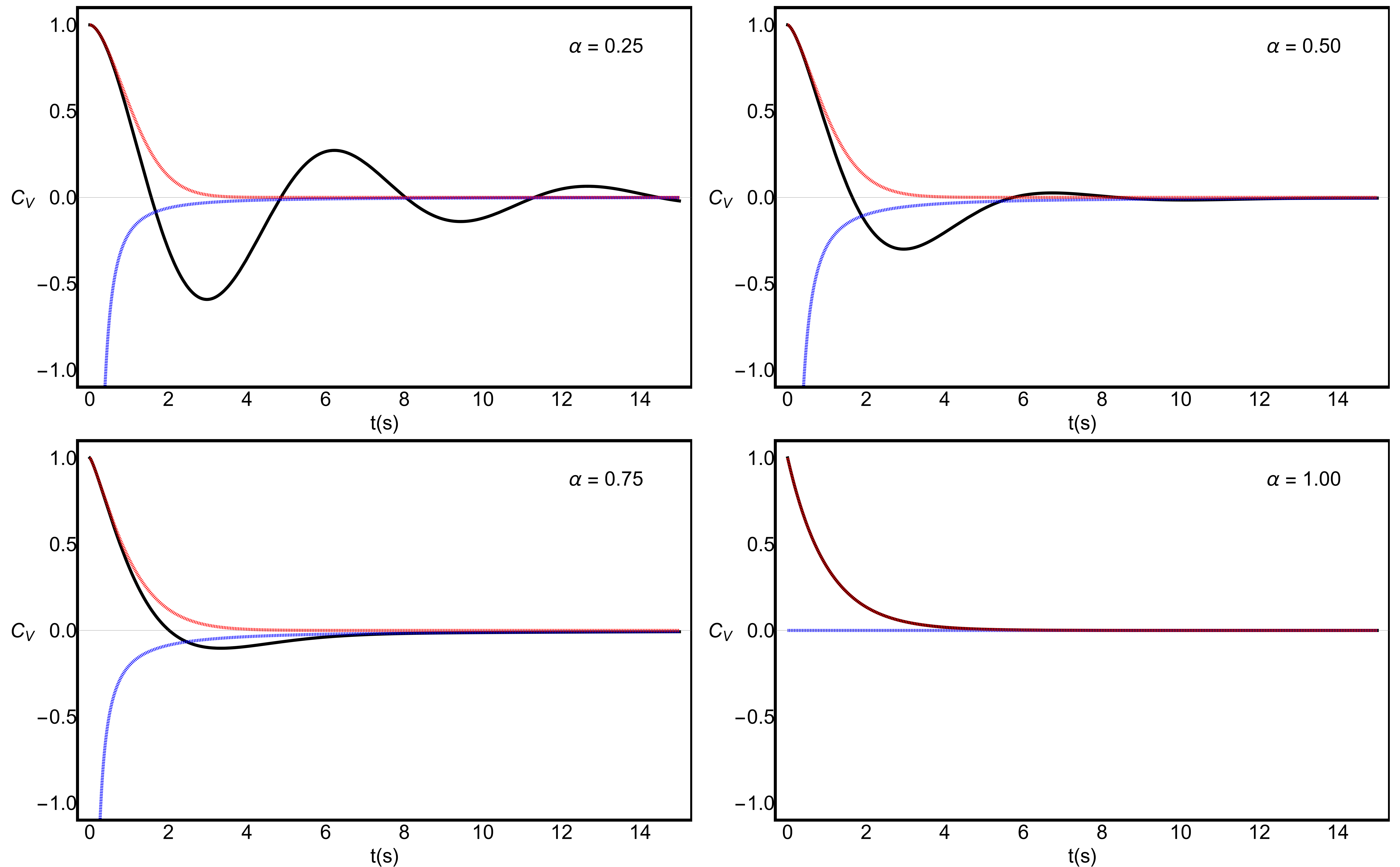}
\caption{Graphical representation of $C_{V}(t)$ vs time $t$ for different scaling exponents $\alpha$ ($0 < \alpha < 1$, ranging 
progressively 
as $0.25, 0.50, 0.75, 1.00$), for fixed 
$k_{B}T=1$ and 
$\gamma_{\alpha}=1$ as described in the text. The figures show the exact relation from Eq.(\ref{eq: vacf anomalo}) (continuous black line) and its asymptotic representations, the stretched exponential from Eq.(\ref{eq: vacf anomalo lim curto}) valid in the short-time limit (dashed red line) and the power law from Eq.(\ref{eq: vacf anomalo lim}) valid in the long-time limit (dashed blue line). Note the different rates of decay for small and large times, where the decay is very fast for short times and very slow for long times.
Here, $\alpha = 1$ yields a special case of the white noise memory in Eq.(\ref{eq: vacf normal lim}).
%The plots show logarithmic scales, where we have chosen the time range $10^{-5}$ $\leq$ $t$ $\leq$ $10^{5}$.
% Here, M-T stands for xx
}
\label{fig: fig2}
\end{figure*}

In what follows, we will consider a 
%slowly decaying 
memory kernel with a power law given by \cite{grebenkov2013, grebenkov2014},
\begin{eqnarray}
    \gamma_{\alpha}(t) = \gamma_{\alpha}\frac{t^{-\alpha}}{\Gamma(1-\alpha)}, \quad 0 < \alpha < 1.
\label{eq: kernel memoria anomala}
\end{eqnarray}
The generalized friction coefficient, $\gamma_{\alpha}$, is independent of time but dependent on the fractional exponent 
%(in units kg $sec^{\alpha − 2}$)%
$\alpha$ and $\Gamma(\cdot)$ stand for the Euler-gamma function. This kernel is known to lead to subdiffusive behavior with the scaling exponent $\alpha$ (for $0 < \alpha < 1$), 
as will be shown in this section. 

Let us now find the relaxation functions $I(t)$, $G(t)$ and $g(t)$ for the memory kernel given in Eq.(\ref{eq: kernel memoria anomala}). 
First, we have that \cite{podlubny1998}
\begin{eqnarray} 
\hat{\gamma}(s) = \mathcal{L}[\gamma(t)](s) = \gamma_{\alpha}s^{\alpha - 1}.
\label{eq: kernel em s}
\end{eqnarray}
So, applying Eq.(\ref{eq: kernel em s}) in (\ref{eq: g trans}), we get
\begin{eqnarray} 
\hat{g}(s) = \frac{1}{s+\gamma_{\alpha}s^{\alpha - 1}}.
\label{eq: g em s}
\end{eqnarray}
The relaxation function $g(t)$ can be found by applying the Laplace inversion of Eq.(\ref{eq: g em s}). To do so, we recall that \cite{podlubny1998},
\begin{eqnarray} 
\mathcal{L} \left[t^{\beta -1}\mathrm{E}_{\alpha,\beta}(\lambda t^{\alpha})\right](s) = \frac{s^{\alpha-\beta}}{s^{\alpha}-\lambda},
\end{eqnarray}
for 
%$\Re(s)>0$, 
$\lambda \in \mathbb{C}$ and $|\lambda s^{-\alpha}| < 1$. So, Eq.(\ref{eq: g em s}) yields 
\begin{eqnarray} 
g(t) = \mathcal{L}^{-1} \left[\hat{g}(s)\right](t) = \mathrm{E}_{2-\alpha, 1}\left(-\gamma_{\alpha} t^{2-\alpha}\right).
\label{eq: g em t}
\end{eqnarray}
The two parameter Mittag-Leffler (M-L) function is introduced and defined by the series expansion \cite{podlubny1998},
\begin{eqnarray}
\mathrm{E}_{\alpha, \beta}(z)=\sum_{k=0}^{\infty} \frac{z^{k}}{\Gamma(\alpha k+\beta)},  
\label{eq: MT definicao}
\end{eqnarray}
where $z \in \mathbb{C}$ and $\alpha>0, \beta>0$. Note that $E_{\alpha, 1}(z)=E_{\alpha}(z)$. 

To proceed, one can use the following identity,
\begin{eqnarray}
\frac{\partial}{\partial t} t^{\beta-1} \mathrm{E}_{\alpha, \beta}\left(\lambda t^{\alpha}\right)=t^{\beta-2} \mathrm{E}_{\alpha, \beta-1}\left(\lambda t^{\alpha}\right),
\end{eqnarray}
and evaluate the remaining relaxation functions in Eqs.(\ref{eq: G trans}) and (\ref{eq: I trans}) from Eq.(\ref{eq: g em s}), 
\begin{eqnarray}
G(t) &=& t \mathrm{E}_{2-\alpha, 2}\left(-\gamma_{\alpha} t^{2-\alpha}\right), \label{eq: G em t} \\
I(t) &=& t^2 \mathrm{E}_{2-\alpha, 3}\left(-\gamma_{\alpha} t^{2-\alpha}\right). \label{eq: I em t}
\end{eqnarray}
The displacement and velocity in Eqs.(\ref{eq: x GLE}) and (\ref{eq: v GLE}) are obtained from Eqs.(\ref{eq: g em t}), (\ref{eq: G em t}), and the initial conditions given by (\ref{eq: termal}). Also from the above relations, the Eqs.(\ref{eq: msd geral})-(\ref{eq: vacf geral}) yields, 
\begin{eqnarray}
\left\langle X^{2}(t)\right\rangle &=&  2 k_{B} T t^2 \mathrm{E}_{2-\alpha, 3}\left(-\gamma_{\alpha} t^{2-\alpha}\right), 
\label{eq: msd anomalo}\\
D(t) &=& k_{B} T t \mathrm{E}_{2-\alpha, 2}\left(-\gamma_{\alpha} t^{2-\alpha}\right), 
\label{qe: D anomalo}\\
C_{V}(t) &=& k_{B} T \mathrm{E}_{2-\alpha, 1}\left(-\gamma_{\alpha} t^{2-\alpha}\right).
\label{eq: vacf anomalo}
\end{eqnarray}
The asymptotic behavior of the relaxation functions relies on the analysis of M-L function \cite{podlubny1998},
\begin{eqnarray}
\mathrm{E}_{\alpha,\beta}\left(-\gamma_{\alpha} t^{\alpha}\right) \simeq 
\begin{cases}
\frac{1}{\Gamma(\beta)}-\frac{\gamma_{\alpha} t^{\alpha}}{\Gamma(\alpha+\beta)},
& \text { as } |\gamma_{\alpha} t^{\alpha}|<<1 \\ 
\frac{t^{-\alpha}}{\gamma_{\alpha} \Gamma(\beta-\alpha)}, 
& \text { as } |\gamma_{\alpha} t^{\alpha}| >>1
\end{cases}
\label{eq: MT limites}
\end{eqnarray}
Introducing these asymptotic expansion in Eqs.(\ref{eq: msd anomalo})-(\ref{eq: vacf anomalo}), we get,
\begin{eqnarray}
\left\langle X^{2}(t)\right\rangle \simeq 2k_{B} T  
\begin{cases}
\frac{t^{2}}{\Gamma(3)}-\frac{\gamma_{\alpha} t^{4-\alpha}}{\Gamma(5-\alpha)},
& \text { as } |\gamma_{\alpha} t^{2-\alpha}|<<1 \\ 
\frac{1}{\gamma_{\alpha}} \frac{t^{\alpha}}{\Gamma(1+\alpha)},
& \text { as } |\gamma_{\alpha} t^{2-\alpha}| >>1
\end{cases}
\label{eq: msd anomalo lim}
\end{eqnarray}
\begin{eqnarray}
D(t) \simeq k_{B} T  
\begin{cases}
\frac{t}{\Gamma(2)}-\frac{\gamma_{\alpha} t^{3-\alpha}}{\Gamma(4-\alpha)},
& \text { as } |\gamma_{\alpha} t^{2-\alpha}|<<1 \\ 
\frac{1}{\gamma_{\alpha}} \frac{t^{\alpha - 1}}{\Gamma(\alpha)},
& \text { as } |\gamma_{\alpha} t^{2-\alpha}| >>1
\end{cases}
\label{eq: D anomalo lim}
\end{eqnarray}
\begin{eqnarray}
C_{V}(t) \simeq k_{B} T  
\begin{cases}
\frac{1}{\Gamma(1)}-\frac{\gamma_{\alpha} t^{2-\alpha}}{\Gamma(3-\alpha)},
& \text { as } |\gamma_{\alpha} t^{2-\alpha}|<<1 \\ 
\frac{1}{\gamma_{\alpha}} \frac{t^{\alpha - 2}}{\Gamma(\alpha-1)},
& \text { as } |\gamma_{\alpha} t^{2-\alpha}| >>1
\end{cases}
\label{eq: vacf anomalo lim}
\end{eqnarray}
Where we observe power law behavior for Eq.(\ref{eq: msd anomalo lim}) in the long-time regime (anomalous diffusion), a slower diffusion (subdiffusive behavior) for $0 < \alpha < 1$, with the generalized diffusion coefficient $D_{\alpha} = k_B T/\gamma_{\alpha}$ \cite{grebenkov2013, grebenkov2014}.

In this paper, we are mainly interested in calculating the NMR signal attenuation by the VACF.
%In this paper, we are mainly interested in calculating the NMR signal attenuation through the VACF.
Graphical representations of the exact VACF, Eq.(\ref{eq: vacf anomalo}), and its asymptotic limits, Eqs.(\ref{eq: vacf anomalo lim}), in the case of thermal initial conditions and different values of the parameter $\alpha$ are given in Fig.(\ref{fig: fig2}), for $0 < \alpha < 1$. We illustrate how the scaling exponent $\alpha$ influences the behavior of the VACF, the other parameters being kept fixed.
%: $k_{B}T=1$ and $\gamma=1$. We stress that $\gamma_{\alpha}$ is a proportionality coefficient  that can depend on the exponent $\alpha$, but is independent of time t. 
As a consequence from Fig.(\ref{fig: fig2}), the VACF interpolates, for intermediate time $t$, between the stretched exponential, expressed from Eq.(\ref{eq: vacf anomalo lim}) in the convergent power series representation,
\begin{eqnarray}
C_{V}(t) \simeq k_{B} T  
\exp \left\{ -\frac{\gamma_{\alpha} t^{2-\alpha}}{\Gamma(3-\alpha)} \right\},
& \text { as } |\gamma_{\alpha} t^{2-\alpha}|<<1~~~~
\label{eq: vacf anomalo lim curto}
\end{eqnarray}
and the negative power law. For small time $t$, the stretched exponential models the very fast decay, whereas for large time $t$ 
it decays with a long negative tail.

\subsection{Normal diffusion}
\label{subsec: Normal diffusion}

The standard Langevin equation for normal diffusion, i.e., without memory, is obtained when $\alpha=1$ in Eq.(\ref{eq: kernel memoria anomala}). In that case, one uses $\gamma_{1}(t) = 2\gamma_1\delta(t)$ to retrieve the instantaneous Stokes force $-\gamma_1 \dot{X}(t)$, where $\delta(t)$ is the Dirac distribution which corresponds to a white noise. So the GLE in Eq.(\ref{eq: lang. geral 2}) yields a special case of the standard Brownian motion and the classical Langevin equation,
\begin{eqnarray}
    \ddot{X}(t) + \gamma_{1}(t) \dot{X}(t) = F(t), \quad t \geq 0.
\end{eqnarray}

The results for the relaxation functions, Eqs.(\ref{eq: g trans})-(\ref{eq: I trans}), arises in a similar manner of Sec. \ref{subsec: Anomalous diffusion}. From that, we get
\begin{eqnarray}
g(t) &=& \mathrm{e}^{-\gamma_{1} t},\\
G(t) &=& (1 - \mathrm{e}^{-\gamma_{1} t})/\gamma_{1},\\
I(t) &=& (\gamma_{1} t - 1 + \mathrm{e}^{-\gamma_{1} t})/\gamma_{1}^2,
\end{eqnarray}
and for Eqs.(\ref{eq: msd geral})-(\ref{eq: vacf geral}), 
\begin{eqnarray}
\left\langle X^{2}(t)\right\rangle &=&  2 k_{B} T (\gamma_{1} t - 1 + \mathrm{e}^{-\gamma_{1} t})/\gamma_{1}^2, 
\label{eq: msd normal}\\
D(t) &=& k_{B} T (1 - \mathrm{e}^{-\gamma_{1} t})/\gamma_{1}, 
\label{qe: D normal}\\
C_{V}(t) &=& k_{B} T \mathrm{e}^{-\gamma_{1} t},
\label{eq: vacf normal}
\end{eqnarray}
where we used that $E_{1,1}(z)=\mathrm{e}^{z}$, $E_{1,2}(z)=\frac{\mathrm{e}^{z}-1}{z}$, and $E_{1,3}(z)=\frac{\mathrm{e}^{z}-1-z}{z^2}$ \cite{podlubny1998}.

The asymptotic behavior from Eqs.(\ref{eq: MT limites}) of the above relations give us,
\begin{eqnarray}
\left\langle X^{2}(t)\right\rangle \simeq 2k_{B} T  
\begin{cases}
\frac{t^{2}}{\Gamma(3)}-\frac{\gamma_{1} t^{3}}{\Gamma(4)},
& \text { as } |\gamma_{1} t|<<1 \\ 
\frac{1}{\gamma_{1}} \frac{t}{\Gamma(2)},
& \text { as } |\gamma_{1} t| >>1
\end{cases}
\label{eq: msd normal lim}
\end{eqnarray}
\begin{eqnarray}
D(t) \simeq k_{B} T  
\begin{cases}
\frac{t}{\Gamma(2)}-\frac{\gamma_{1} t^{2}}{\Gamma(3)},
& \text { as } |\gamma_{1} t|<<1 \\ 
\frac{1}{\gamma_{1}},
& \text { as } |\gamma_{1} t| >>1
\end{cases}
\label{eq: D normal lim}
\end{eqnarray}
\begin{eqnarray}
C_{V}(t) \simeq k_{B} T  
\begin{cases}
\frac{1}{\Gamma(1)}-\frac{\gamma_{1} t}{\Gamma(2)},
& \text { as } |\gamma_{1} t|<<1 \\ 
0,
& \text { as } |\gamma_{1} t| >>1
\end{cases}
\label{eq: vacf normal lim}
\end{eqnarray}
Where we observe normal diffusive behavior for Eq.(\ref{eq: msd normal lim}) in the long-time regime, as we expected.%\\

\subsection{NMR signal attenuation}
\label{subsec: NMR signal attenuation}

We 
%now turn the attention and 
can now use the results from previous section to evaluate the NMR signal attenuation due do diffusion in both anomalous and normal cases. 
%In this paper, we are mainly interested in calculate this effect trough the VACF given by 
So, from Eq.(\ref{eq: fase gaussiana numa direcao}) and (\ref{eq: vacf anomalo}), we have
\begin{eqnarray}
\left\langle \frac{\Phi^{2}}{2}\right\rangle = 
\gamma_{n}^{2}g^{2} \mathrm{\textbf{I}},
\label{eq: fase anomala}
\end{eqnarray}
with the following integral to be solved,
\begin{eqnarray}
\mathrm{\textbf{I}} &=&
k_{B} T  \nonumber \\ 
&&
%\times
\int_{0}^{T_E} 
\int_{0}^{T_E} 
t_{1} t_{2}
\mathrm{E}_{2-\alpha, 1}\left(-\gamma_{\alpha} |t_{1}-t_{2}|^{2-\alpha}\right)
\mathrm{d} t_{1} 
\mathrm{d} t_{2}.~~~~~~
\label{eq: inte anomala}
\end{eqnarray}
The Eq.(\ref{eq: inte anomala}) may be integrated 
(see, e.g, \cite{podlubny1998}),
%(see Appendix), 
and one obtains the exacts results for the effective gradient in Eq.(\ref{eq: gradiente efetivo}),
%\begin{widetext}
\begin{eqnarray}
\mathrm{\textbf{I}}
=
k_{B} T  
\begin{cases}
\tau^4 [2\mathrm{E}_{2-\alpha, 5}\left(-\gamma_{\alpha} \tau^{2-\alpha}\right)],
& t = \tau \\ 
\tau^{4} 
[2^{4} \mathrm{E}_{2-\alpha, 5}\left(-\gamma_{\alpha} (2\tau)^{2-\alpha}\right) 
\\ \quad - 2^{2}
\mathrm{E}_{2-\alpha, 5}\left(-\gamma_{\alpha} \tau^{2-\alpha}\right)],
& t = 2\tau
\end{cases}
\label{eq: inte anomala resolvida}
\end{eqnarray}
%\end{widetext}
The exact attenuation in Eq.(\ref{eq: atenuacao da fase gaussiana}) is then given by Eqs.(\ref{eq: inte anomala resolvida}) and (\ref{eq: fase anomala}). In the long-time limit yields,
\begin{eqnarray}
E \simeq
\begin{cases}
\exp \left\{-
    \frac{2}{\Gamma(\alpha + 3)}
    \gamma_{n}^{2}g^{2}D_{\alpha}
    \tau^{\alpha+2}
    \right\},
& t = \tau \\ 
\exp \left\{-
    \frac{\left(1-1/2^{\alpha}\right)}{\Gamma(\alpha + 3)}
    \gamma_{n}^{2}g^{2}D_{\alpha}
    T_E^{\alpha+2}
    \right\},
& t = 2\tau
\end{cases}
\label{eq: sinal anomalo}
\end{eqnarray}
Here, in Eq.(\ref{eq: sinal anomalo}) we see that when the system is driven by the memory kernel in Eq.(\ref{eq: kernel memoria anomala}), the fractional exponent $\alpha$ plays a role in the NMR signal attenuation. Note that the relation (\ref{eq: sinal anomalo}) agrees with the one obtained by K{\"a}rger et al. approach \cite{karger1988} for the case expressed by Eq.(\ref{eq: kernel memoria anomala}).

At last, for normal diffusion ($\alpha$ = 1) in Eq.(\ref{eq: sinal anomalo}), 
\begin{eqnarray}
E =  
\begin{cases}
\exp \left\{-
    \frac{1}{3} \gamma_{n}^2 g^2 D \tau^3
    \right\},
& t = \tau \\ 
\exp \left\{-
    \frac{1}{12} \gamma_{n}^2 g^2 D T_{E}^3
    \right\},
& t = 2\tau
\end{cases}
\label{eq: sinal normal}
\end{eqnarray}
one retrieves the Hanh echo \cite{hahn1950} expected result of the NMR signal attenuation due to normal diffusion as in Eq.(\ref{eq: hahn echo}).

\section{Conclusions Remarks}
\label{sec: Conclusions Remarks}

In the present work, with the application of a magnetic field gradient, it was possible to encode the nuclei trajectories. We investigate the underlying stochastic processes using a phenomenological description provide by a generalized Langevin equation (GLE). 
While in simple systems the memory friction kernel yields a white noise, in complex environments the picture is different, exhibiting some form of memory. We have chosen a memory kernel that decays as a power law with the scaling exponent $\alpha$, which leads to anomalous behavior. Namely, the relaxation functions are of a power law type and were expressed in terms of generalized Mittag-Leffler functions and their derivatives. The asymptotic behavior of these quantities was obtained and reveals, in the long-time regime, subdiffusion behavior for $0 < \alpha < 1$.

We have shown how one can apply the results from this approach to evaluate the Nuclear
Magnetic Resonance (NMR) signal attenuation due to diffusion
%, in the long time limit, 
as a function of the velocity autocorrelation function (VACF), a relevant quantity that provides details of molecular dynamics and that can be measured in the laboratory \cite{stepisnik2000}.
%The VACF was obtained using the Langevin stochastic approach 
%in two distinct dynamic cases, namely, the standard and a fractional Langevin equation (FLE). 
%In both cases, 
Here, we consider acquisitions made with a steady gradient and within the framework of the Gaussian phase approximation (GPA). We emphasize that other gradient temporal profiles and pulse sequences could also be used but the above case capture all the essential features of the gradient NMR experiment.

On the one hand, we demonstrate the advantages of using the GLE approach in theoretical studies of anomalous diffusion. On the other hand, the mathematical description allows one to evaluate and properly interpret, in the long-time diffusion regime, the expression for diffusion-based NMR experiments. The choice of a memory kernel that decays as a power law allows one to perform the calculation analytically, illustrate the emergence of fractional order exponential decay in the NMR signal amplitude and retrieve the classic result from Hanh echo \cite{hahn1950}. Finally, we conclude that the solutions may also be extended to describe the dynamics of a broad class of systems, depending on the memory kernel $\gamma(t)$ \cite{grebenkov2013, grebenkov2014}. So we expect the obtained results from this framework to be useful for a better description of experimental data.

\section*{Acknowledgments} 
FPA gratefully acknowledges financial support from the Brazilian agency CNPq (Grant No. 142764/2020-5). DOSP acknowledges the Brazilian funding agencies CNPq (Grant No. 307028/2019-4), FAPESP (Grant No. 2017/03727-0), and the Brazilian National Institute of Science and Technology of Quantum Information (INCT/IQ). 
FFP acknowledges the Brazilian agency CNPq (Grant No. 425346/2018-8).

\bibliography{p1-fpa}
\end{document}